\documentclass[a4paper,11pt]{article}

\usepackage{amsfonts,amssymb,amsmath,amsthm,dsfont,xfrac,xspace}
\usepackage{fullpage}
\usepackage{graphicx}
\usepackage{subfig}
\usepackage{float}
\usepackage{cite}
\usepackage{hyperref}
\usepackage{algorithmic}
\graphicspath{{./},{fig/}}
\makeatletter
\let\NAT@parse\undefined
\makeatother
\usepackage[sort&compress, numbers]{natbib}
\usepackage{hyperref}

\usepackage{mathtools,bbm}
\usepackage{cuted}
\usepackage{multicol}
\usepackage{stfloats}

\newsavebox{\ieeealgbox}

\renewcommand{\thelem}{\the \numexpr (\value{thm}+1) \relax.\arabic{lem}}

\title{Frequency-Domain Time-Reversal Precoding in Wideband MISO OFDM Communication Systems} 

\author{Trung-Hien~Nguyen\thanks{T.-H.~Nguyen,  J.-F.~Determe, M.~Van~Eeckhaute, P.~De~Doncker, and F.~Horlin are with OPERA department, Universit\'e libre de Bruxelles (ULB), 1050 Brussels, Belgium. E-mail: trung-hien.nguyen@ulb.ac.be} , Jean-Fran\c cois~Determe$^*$, Mathieu~Van~Eeckhaute$^*$, \\ 
J\'er\^ome~Louveaux\thanks{J.~Louveaux is with ICTEAM institute, Universit\'e catholique de Louvain (UCL), 1348 Louvain-la-Neuve, Belgium.} , Philippe~De~Doncker$^*$, and Fran\c cois~Horlin$^*$  } 

\begin{document}
\maketitle

\begin{abstract}
	Time reversal (TR) recently emerged as an interesting communication technology capable of providing a good spatio-temporal signal focusing effect. New generations of large-bandwidth devices with reduced cost leverage the use of TR wideband communication systems. While TR is usually implemented in the time domain, the same benefit can be obtained in an orthogonal frequency division multiplexing (OFDM) system by precoding the information in the frequency domain. Besides using multiple antennas, the focusing effect of TR also comes from the use of a high rate back-off factor (BOF), which is the signal up-sampling (or down-sampling) rate in the original time-domain TR precoding. However, a frequency-domain TR precoding in the literature has only considered BOF of one, which does not fully exploit the focusing property of TR. In this paper, we discuss how to properly implement different BOFs using frequency-domain TR precoding in the OFDM system. Moreover, we demonstrate that increasing the BOF and/or the number of transmit antennas significantly improves the focusing gain at the intended position. In contrast, the unintended positions receive less useful power. Furthermore, closed-form approximations of the mean-square-errors (MSEs) of equalized received signals at either intended or unintended positions are derived, expressing the focusing gain as a function of the BOF and the number of antennas. Numerical simulations with multi-path channels are carried out to validate the MSE expressions.
\end{abstract}


\section{Introduction}

Time-reversal (TR) with a sufficiently high rate back-off factor (BOF) has recently gained much attention as it can combine the energy from different multipath components to create a spatio-temporal focusing effect~\cite{ChenSPL13},~\cite{EmamiACSSC04}. TR precoding can be implemented in either time-domain (TD) or frequency-domain (FD). The TD/TR precoding has been well studied in the literature~\cite{EmamiACSSC04}. The focusing effect of the TD/TR precoding comes from the use of a high BOF, which is defined as the signal up-sampling rate at the transmitter before the precoding process (or the respective signal down-sampling rate at the receiver before the equalization). Because of the spatio-temporal focusing effect of TD/TR precoding systems, a one-tap equalizer is generally sufficient in such systems enabling a complexity reduction at the receiver. % ,~\cite{WangJSAC11}

Orthogonal frequency-division multiplexing (OFDM) is an efficient technology for new generation wireless systems, thanks to its simple equalization of the frequency-selective channel. The FD/TR precoding combined with OFDM has been shown to be a simple and efficient technology~\cite{DuboisJWCN13}. However, the FD/TR precoding in the literature did not consider different BOFs and the impact of this parameter on the focusing effect, hence the advantage of TR was not exploited. For instance, in~\cite{DuboisJWCN13}, the authors have derived the closed-form expression of the bit-error-rate (BER) of FD/TR precoding multiple-input single-output (MISO) OFDM systems with implicitly a BOF of one.

In this paper, we present a proper way to perform TR precoding in the FD with different BOFs and verify the focusing gain of this FD/TR precoding in the MISO OFDM system. \emph{The focusing gain is defined by the signal-to-noise ratio (SNR) gain necessary to keep the mean-square-error (MSE) of the received signals at a fixed value}. Our main contributions can be summarized as follows: 
\begin{itemize}
\item We investigate for the first time the equivalent TR precoding method in the FD for MISO OFDM systems and assess the focusing gain with different BOFs.
\item We derive closed-form MSE approximations in the FD/TR precoding MISO OFDM systems at the intended and unintended positions. Similarly to the TD/TR precoding, the MSE of the received symbols at the intended position is shown to be much less than that at the unintended position, confirming the focusing gain of the FD/TR precoding. In order to obtain the MSE approximation at the unintended position, we derive the probability density function (PDF) of the modulus of the sum of products of zero-mean complex Gaussian random variables (RVs) with arbitrary variances.
\item We analyze the asymptotic behaviors of the MSE expressions to understand in a simple manner the tendencies evolution of the MSE at either intended or unintended positions. Finally, numerical simulations are carried out to validate our analysis.
\end{itemize}

The remainder of this paper is organized as follows: we present a brief overview on the TR technique in Section II; the TR-based system model and the proper way to perform TR precoding in the FD are introduced in Section III; we carry out theoretical performance analyses in Section IV; the asymptotic analysis is conducted in Section V; we present the analytical and numerical results in Section VI; finally, the paper is concluded in Section VII.

\emph{Notation}: the underlined lower-case and upper-case letters denote column vectors of TD and FD variables, respectively. Double-underlined upper-case letter corresponds to a matrix; $\underline{\underline{I}}_N$ is the $N \times N$ identity matrix; $\underline{\underline{0}}_N$ is the $N \times N$ zero matrix; $\underline{\underline{F}}_Q$ is the $Q \times Q$ Fourier matrix; $| \cdot |$, $|| \cdot ||$, $(\cdot)^*$, $(\cdot)^T$, $(\cdot)^H$ are the absolute, norm, complex conjugate, transpose and  Hermitian transpose operators, respectively; $\rm{tr} \{ \cdot \}$ and ${\mathbb{E}}[\cdot]$ are the trace and expectation operators, respectively; $x!$ is the factorial of a positive integer $x$.

\section{Time Reversal History}
\label{secHis}

The TR technique was invented to focus the wave energy into one position over the space~\cite{FinkTAP02}. Actually the TR has been first proposed in 1950's in~\cite{BogertTCS57}, where it was applied to alleviate the delay distortion in a picture transmission system. TR was then investigated in the digital communications in the 1960's, and it has been shown to be the optimal solution of a minimum-transmission-lost constrained optimization problem in the digital communication systems~\cite{AmorosoTCT66}.

Until 1990's the TR has gained new interests with the application in the ultrasonic and acoustics systems~\cite{FinkUS89}, \cite{FinkTAP02}. Since then TR has been applied in a lot of applications based on its spatio-temporal focusing effect. For example, the TR imaging technique has been proposed to identify multiple targets within a specific area in combination with the multiple signal classification (MUSIC) algorithm~\cite{DevaneyTAP05}. Target detection using the TR-based antenna array has been investigated in multipath scattering environments~\cite{JinTSP09}. Recently, TR has been claimed to be a promising candidate for the 5G communication systems, where massive multipaths (one of the major challenges in modern wireless communications) are exploited to enhance the communication link instead of trying to remove them~\cite{ChenSPM16}. TR is also shown to be an interesting technology to provide physical layer security~\cite{XuTVT18}, centimeter-accuracy indoor localization~\cite{WuTVT15, NguyenWiMob18}, event detection through a wall~\cite{XuIoT17}, monitoring of vital signs~\cite{ChenTBE17}. It is worth noticing that almost all TR-based applications referred here are in the TD. The FD/TR applications, however, did not consider carefully the equivalent TR version as originally invented, i.e., that for which BOF is larger than one. 

Due to the fact that OFDM modulation is used widely nowadays - especially in the 5G networks, which leverages the use of FD precoding techniques for complexity reduction - the aforementioned TD/TR precoding makes it difficult to integrate into the system. In what follows, we discuss in detail the corresponding FD/TR precoding.

\section{System Model}
\label{sec2}

\begin{figure}%[!htbp]%[!t]
\centering
\includegraphics[width=0.68\textwidth]{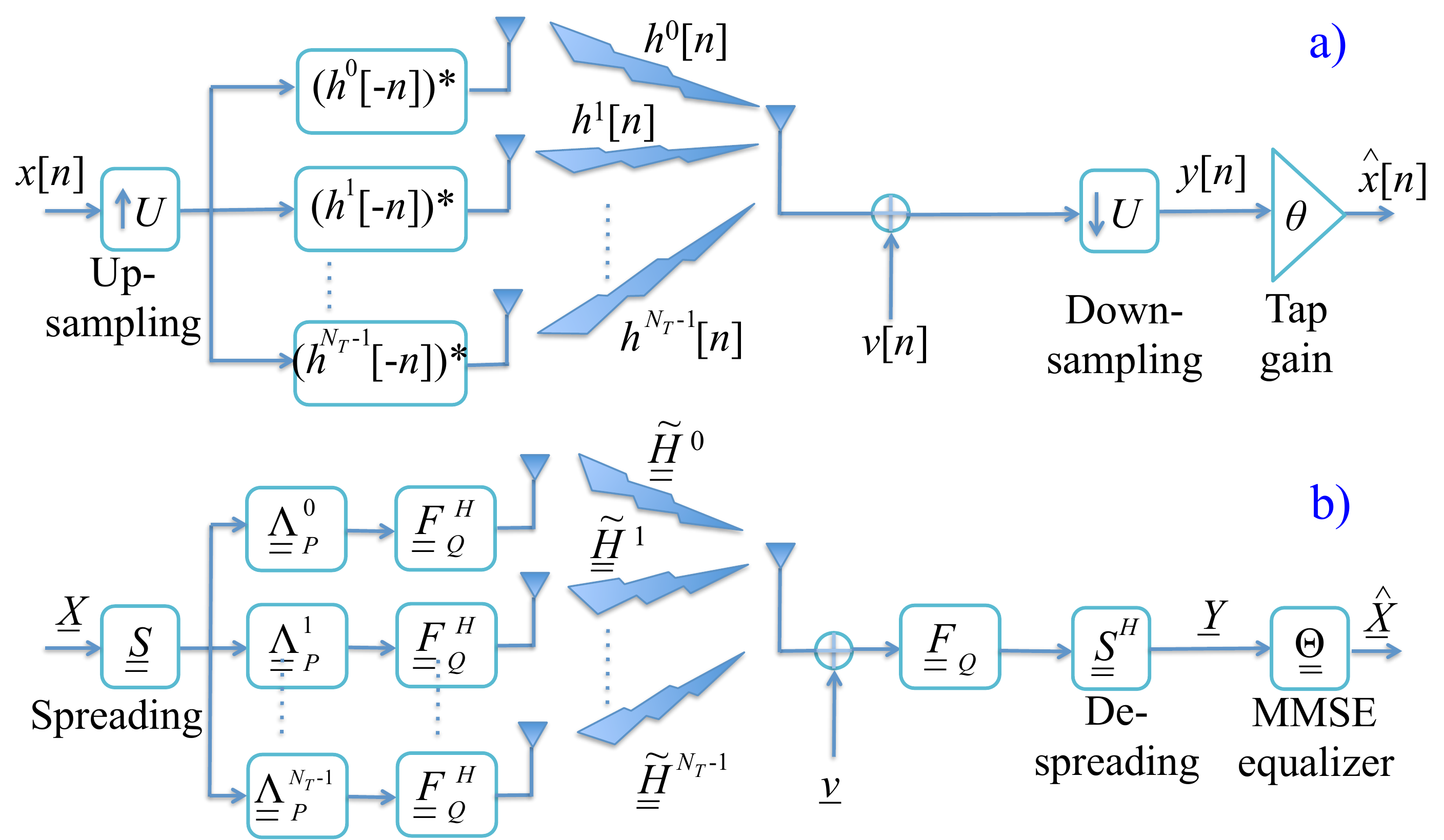}
\caption{Schematics of (a) the TD/TR precoding communication system~\cite{EmamiACSSC04} and (b) the corresponding FD/TR precoding OFDM system.}
\label{fig1}
\end{figure}

The TD/TR precoding MISO system is illustrated in Fig.~\ref{fig1}(a)~\cite{EmamiACSSC04}, where the TD symbol sequence $x[n]$ is firstly up-sampled by a BOF $U$ and repeated on $N_T$ transmit antenna branches. The signals are then pre-filtered by the TR precoder, $(h^k [-n])^*$, of the associated $k$-th channel impulse response (CIR), $h^k [n]$ before being sent over antennas. At the receiver side, the signal corrupted by additive white Gaussian noise (AWGN) $v[n]$ is down-sampled by $U$. It is then equalized by a tap gain $\theta$, which can be designed based on the minimum mean square error (MMSE) criterion. It has been shown in~\cite{EmamiACSSC04} that by using more antennas and/or increasing the BOF, the spatio-temporal focusing gain is improved accordingly, delivering a better bit-error-ratio (BER)/MSE performance at the cost of the signal rate reduction. Note that, the TD/TR precoding system requires a wide-bandwidth analog-to-digital converter (ADC) at the receiver. However, the sampling rate can be kept low.

In the literature, the performance of FD/TR precoding OFDM system was only assessed for a varying number of antennas~\cite{DuboisJWCN13}, while the use of different BOFs has not been studied yet. We discuss here a proper way to assign the data symbols onto OFDM subcarriers in FD/TR precoding system, as illustrated in Fig.~\ref{fig1}(b). The number of subcarriers of one OFDM symbol is $Q$. For simplicity, we consider that a single OFDM symbol is sent over the TR precoding MISO OFDM system. 
We consider a transmit data vector $\underline{X}$ composed of $N$ symbols $X_n$ (for $n=0,1,...,N-1$ with $N=Q/U$), i.e., $\underline{X} = [X_0 \ ... \ X_{N-1}]^T$. The symbols $\{ X_n \}$ are assumed to be independent zero-mean random variables with variance $\mathbb{E}[|X_n|^2] = \sigma_X^2$. Without loss of generality, a normalized constellation is considered, i.e., $\sigma_X^2=1$. 

The data symbols $\underline{X}$ are then spread by the matrix $\underline{\underline{S}}$ of size $Q \times N$. The $Q \times N$ matrix $\underline{\underline{S}}$ is the concatenation of $U$ independent $N \times N$ diagonal matrices, whose diagonal values are identically and independently distributed and taken from the set $\{ \pm 1 \}$. The spreading matrix is normalized by $\sqrt{U}$ in order to get ${\underline{\underline S} ^H}\underline{\underline S}  = {\underline{\underline I} _N}$. Thanks to the spreading matrix, the BOF discussed in the original TD/TR precoding is properly introduced. 
The idea behind this spreading comes from the fact that up-sampling a signal in the TD is equivalent to the repetition and shifting of its spectrum in the FD. Furthermore, a randomized spreading code is used to avoid assigning the same data symbol on different OFDM subcarriers, which causes a high peak-to-average-power ratio (PAPR)~\cite{SpethTC99}. 

After spreading, the signal is repeated on $N_T$ branches corresponding to transmit antennas and pre-coded by a matrix $\underline{\underline{\Lambda}} _P^k$ on each branch $k$. Defining ${\underline H ^k} := {[H_0^k \ {\rm{  }}H_1^k \ {\rm{ }}... \ {\rm{ }}H_{Q - 1}^k]^T}$ as the channel frequency response (CFR) associated with the $k-$th antenna and assuming it is normalized such that $|| \underline{H} ^k || = 1$, then $\underline{\underline{\Lambda}} _P^k$ is the diagonal matrix, whose diagonal elements are $(H_q ^k)^*$ (for $q = 0, 1, ..., Q-1$). The FD/TR precoded signals are transformed to TD signals by using an inverse fast Fourier transform (IFFT) operator. 
The signal is then made cyclic by adding/removing a cyclic prefix (CP) and propagated over the channel, which is mathematically equivalent to the multiplication with the  $Q \times Q$ circulant matrix ${{{\underline{\underline {\widetilde H}} }^k}}$ of the $k-$th CIR. The matrix ${{{\underline{\underline {\widetilde H}} }^k}}$ can be factorized as ${\underline{\underline {\widetilde H}} ^k} = \underline{\underline F} _Q^H \cdot \underline{\underline \Lambda } _H^k \cdot {\underline{\underline F} _Q}$, where $\underline{\underline \Lambda } _H^k$ is the diagonal matrix, whose diagonal elements are the elements of vector $\underline{H} ^k$. It is worth noticing that $\underline{\underline \Lambda } _H^k$ is associated to an arbitrary position, which may or may not be the intended position, i.e., that for which the precoding matrix $\underline{\underline{\Lambda}} _P^k$ is computed. At the receiver, the reversed operations are carried out. Note that, a wide-bandwidth ADC is required and a high sampling rate is needed as in conventional OFDM systems. However, the de-spreading operation reduces the sample rate to the symbol rate, thereby enabling the receiver to work at a \emph{low-rate processing}. Assuming the time and frequency synchronization is perfect, the received signal after de-spreading is given by

\begin{equation}
\underline Y  = {\underline{\underline S} ^H} \cdot {\underline{\underline F} _Q} \cdot \left( {\sum\limits_{k = 0}^{{N_T} - 1} {{{\underline{\underline {\widetilde H}} }^k} \cdot \underline{\underline F} _Q^H \cdot \underline{\underline \Lambda } _P^k} } \right) \cdot \underline{\underline S}  \cdot \underline X  + \underline V ' ,
\label{eq1}
\end{equation}

\noindent where $\underline V ' = {\underline{\underline S} ^H}{\underline{\underline F} _Q}\underline v $ is the equivalent FD circularly symmetric complex AWGN of the TD AWGN $\underline{v}$. We assume that the signal $X_n$ and noise $V'_n$ are independent of each other. We define the noise auto-correlation matrix as ${\underline{\underline R} _{vv}}: = {\mathbb{E}}\left[ {\underline v  \cdot  {{\underline v }^H}} \right] = \sigma _v^2{\underline{\underline I} _Q}$, where $\sigma _v^2$ is the variance of $\underline{v}$. Based on the definition of the spreading matrix, we can deduce ${\underline{\underline R} _{V'V'}} = \sigma _v^2{\underline{\underline I} _N}$. After some manipulations, \eqref{eq1} can be rewritten as $\underline Y  = \underline{\underline G} \cdot \underline X  + \underline V '$, where $\underline{\underline G}  = {\underline{\underline S} ^H} \cdot \left( {\sum\nolimits_{k = 0}^{{N_T} - 1} {\underline{\underline \Lambda } _H^k \cdot \underline{\underline \Lambda } _P^k} } \right) \cdot \underline{\underline S} $. 

Similarly to the TD/TR system, we use a one-tap MMSE equalizer after de-spreading in order to recover the transmitted signal. By multiplying $\underline{Y}$ with the MMSE equalizer matrix, given by $\underline{\underline \Theta}  = {\left( {{{\underline{\underline G} }^H} \cdot \underline{\underline G}  + {\gamma ^{ - 1}}{{\underline{\underline I} }_N}} \right)^{ - 1}} \cdot {\underline{\underline G} ^H}$ (in which $\gamma : = {{\sigma _X^2} \mathord{\left/ {\vphantom {{\sigma _X^2} {\sigma _v^2}}} \right. \kern-\nulldelimiterspace} {\sigma _v^2}}$ is the definition of the SNR), we obtain the estimate $\underline{\widehat X}$ of the input signal vector. It is worth noticing that if the TR precoding is matched to the channel, $\underline{\underline{\Theta}}$ is a real-valued diagonal matrix, leading to a \emph{low-complexity equalizer} at the receiver.

\section{Performance Assessment}
\label{sec3}

By defining $\underline \varepsilon  : = \underline X  - \underline {\widehat X} $ and ${\underline{\underline R} _{\varepsilon \varepsilon }}: = {\mathbb{E}}\left[ {\underline \varepsilon \cdot  {{\underline \varepsilon  }^H}} \right]$ and after some manipulations, the MSE of the equalized received symbol can be derived as follows%~\cite{KleinTVT96}

\begin{equation}
MSE = {\rm{tr}}\left\{ {{{\underline{\underline R} }_{\varepsilon \varepsilon }}} \right\} = \sigma _v^2 \cdot {\rm{tr}}{\left\{ {{\mathbb{E}}\left[ {{{\left( {{{\underline{\underline G} }^H} \cdot \underline{\underline G}  + {\gamma ^{ - 1}}{{\underline{\underline I} }_N}} \right)}^{ - 1}}} \right]} \right\} }. %^*} 
\label{eq4}
\end{equation}

In order to assess the focusing gain of TR, we compare the normalized MSEs (NMSEs) (defined as $NMSE: = {{MSE} \mathord{\left/ {\vphantom {{MSE} {\left( {N\sigma _X^2} \right)}}} \right. \kern-\nulldelimiterspace} {\left( {N\sigma _X^2} \right)}}$) of the received signal at both the intended and unintended communication positions. Due to the fact that $\underline{\underline{G}}$ is a diagonal matrix, the NMSE is given by

\begin{equation}
NMSE = \frac{1}{N}\sum\limits_{n = 0}^{N - 1} {{\mathbb{E}}\left[ {\frac{{{\gamma ^{ - 1}}}}{{\frac{1}{{{U^2}}}{{\left| {{K_n}} \right|}^2} + {\gamma ^{ - 1}}}}} \right]} ,
\label{eq5}
\end{equation}

\noindent where $K_n$ is a random variable (RV) depending on the channel realization. The RV ${K_n} = \sum\nolimits_{k = 0}^{{N_T} - 1} {\sum\nolimits_{u = 0}^{U - 1} {{{\left| {H_{n + uN}^k} \right|}^2}} } $ and ${K_n} = \sum\nolimits_{k = 0}^{{N_T} - 1} {\sum\nolimits_{u = 0}^{U - 1} {\overline H _{n + uN}^k \cdot {{\left( {H_{n + uN}^k} \right)}^*}} } $ at the intended and unintended positions, respectively, in which $\overline{H} _n^k$ is the $n$-th component of the unintended position CFR associated with the $k$-th transmit antenna. We assume that the channels at the intended and unintended postions are spatially independent.

We further assume that the CIRs between each transmit antenna and the receive antenna comprise no more than \emph{L} taps and are independent. The variance of the \emph{l}-th CIR tap is $\sigma _{{h_l}}^2 = {\mathbb{E}}\left[ {{{\left| {{h_l}} \right|}^2}} \right]$. Based on these assumptions that makes sure the CFR components are uniformly spaced over the bandwidth and thanks to the design of spreading matrix $\underline{\underline{S}}$ as described in Section~\ref{sec2}, the RV $K_n$ in (\ref{eq5}) is constructed by the sum of weakly correlated complex RVs. When the BOF value is small or moderate, RVs $H_{n+uN}^k$ and $H_{n+(u+1)N}^k$ (for $\forall n \in [0,N-1]$, $\forall k \in [0,N_T-1]$ and $\forall u \in [0,U-2]$) can be considered to be independent from one to another and identically distributed so that the NMSE in~\eqref{eq5} can be approximated by

\begin{equation}
NMSE \approx {\mathbb{E}}\left[ {\frac{{{\gamma ^{ - 1}}}}{{\frac{1}{{{U^2}}}{{\left| {{K_n}} \right|}^2} + {\gamma ^{ - 1}}}}} \right] .
\label{eq5b}
\end{equation}

\noindent \emph{- \underline{At the intended position:}} 

The RV ${K_n}$ has the PDF ${f_Z}\left( z \right) = {{{z^{M - 1}}{{\rm{e}}^{ - z}}} \mathord{\left/ {\vphantom {{{z^{M - 1}}{{\rm{e}}^{ - z}}} {\left( {M - 1} \right)!}}} \right. \kern-\nulldelimiterspace} {\left( {M - 1} \right)!}}$~\cite{PapoulisBook91}, where $M = U{N_T}$. Applying this PDF to \eqref{eq5b}, we obtain a closed-form NMSE expression% as follows

\begin{equation}
NMSE = \frac{{{\gamma ^{ - 1}}}}{{\left( {M - 1} \right)!}}\int\limits_0^\infty  {\frac{{{z^{M - 1}}}}{{{{{z^2}} \mathord{\left/
 {\vphantom {{{z^2}} {{U^2}}}} \right. \kern-\nulldelimiterspace} {{U^2}}} + {\gamma ^{ - 1}}}}{{\rm{e}}^{ - z}}dz} .
\label{eq5a}
\end{equation}

Although \eqref{eq5a} can be numerically computed, it is still difficult to analyze the asymptotic behaviors. We therefore approximate \eqref{eq5a} as in the Appendix~A to yield a tractable NMSE approximation at the intended position

    \begin{align}
	  NMSE \approx \frac{1}{{\left( {U{N_T} - 1} \right)!}} & \left( {{\Gamma _{low}}\left( {U{N_T},U{\gamma ^{ - 1/2}}} \right) - \frac{1}{{{U^2}{\gamma ^{ - 1}}}}{\Gamma _{low}}\left( {U{N_T} + 2,U{\gamma ^{ - 1/2}}} \right)   } \right.     \nonumber \\
	  &  {\rm{ \ \ + \ }}\frac{1}{{{U^4}{\gamma ^{ - 2}}}}{\Gamma _{low}}\left( {U{N_T} + 4,U{\gamma ^{ - 1/2}}} \right)  + {U^2}{\gamma ^{ - 1}}{\Gamma _{up}}\left( {U{N_T} - 2,U{\gamma ^{ - 1/2}}} \right)   \nonumber \\
	  &  \left. { \ \   - \ {U^4}{\gamma ^{ - 2}}{\Gamma _{up}}\left( {U{N_T} - 4,U{\gamma ^{ - 1/2}}} \right)} \right) .
       \label{eq6}
    \end{align}

\noindent in which $\Gamma_{low}$ and $\Gamma_{up}$ are the lower and upper incomplete Gamma functions, respectively, and defined by

\begin{equation}
{\Gamma _{low}}\left( {a,t} \right) = \int_0^t {{x^{a - 1}}{{\rm{e}}^{ - x}}dx} ,
\label{eq7}
\end{equation}

\begin{equation}
{\Gamma _{up}}\left( {a,t} \right) = \int_t^\infty  {{x^{a - 1}}{{\rm{e}}^{ - x}}dx} ,
\label{eq8}
\end{equation}

\noindent such that ${\Gamma _{low}}\left( {a,t} \right) + {\Gamma _{up}}\left( {a,t} \right) = \Gamma \left( a \right)$ is the Gamma function.

\noindent \emph{- \underline{At the unintended position:}} 

In order to derive a closed-form NMSE approximation, we have to know the PDF of the RV ${\left| {{K_n}} \right|}$. This PDF is derived in the following theorem.

\noindent \textbf{Theorem:} Let a RV $Z_m = Y_{1,m} \cdot Y_{2,m}$, where $Y_{1,m}$ and $Y_{2,m}$ are statistically independent and identically distributed (i.i.d) complex Gaussian RVs of zero-mean and variance $\sigma_{Y_1}^2$ and $\sigma_{Y_2}^2$, respectively, i.e., $Y_{1,m} \sim \mathcal{CN}(0, \,\sigma_{Y_1}^{2})\,$, $Y_{2,m} \sim \mathcal{CN}(0, \,\sigma_{Y_2}^{2})\,$. Considering $ Z = \sum\nolimits_{m = 0}^{M - 1} {{Z_m}} $, then the modulus of $Z$, ${R} = \left| Z \right|$, has the following PDF
% , in which $Z_0, \ Z_1, \ ..., \ Z_{M-1}$ are statistically independent from one to another with the same distribution

\begin{equation}
{f_{{R}}}\left( r \right) = \frac{{4{r^M}}}{{\Gamma \left( M \right) \cdot {{\left( {{\sigma _{Y_1}}{\sigma _{Y_2}}} \right)}^{M + 1}}}}{ \mathbb{K}_{M - 1}}\left( {\frac{{2r}}{{{\sigma _{Y_1}}{\sigma _{Y_2}}}}} \right) ,
\label{eq8pdf}
\end{equation}

\noindent where $\mathbb{K}_{M} ( \cdot )$ is the $M$-th order modified Bessel function of the second kind.

\noindent \emph{Proof:} See Appendix~B.

In our study, we consider that each TD channel realization has a power normalized to one, which also induces that the variances of RVs $H_n^k$ and $\overline{H} _n^k$ are equal to one. Substituting the result of the theorem into \eqref{eq5b}, we obtain the following closed-form NMSE expression at the unintended position

\begin{equation}
NMSE = \frac{{4{\gamma ^{ - 1}}}}{{\Gamma \left( M \right)}}\int\limits_0^\infty  {\frac{{{z^M}}}{{{{{z^2}} \mathord{\left/
 {\vphantom {{{z^2}} {{U^2}}}} \right. \kern-\nulldelimiterspace} {{U^2}}} + {\gamma ^{ - 1}}}}{\mathbb{K}_{M - 1}}\left( {2z} \right)dz} .
\label{eq8b}
\end{equation}

In order to get a tractable NMSE expression to further evaluate the asymptotic behaviors, we apply the series expansion of the modified Bessel function of the second kind~[18,~eq.~(24)]

\begin{equation}
{\mathbb{K}_M}\left( x \right) \approx \sum\limits_{q = 0}^D  {\sum\limits_{l = q}^D {\Psi \left( {M,l,q} \right)} }   \cdot {{\rm{e}}^{ - x}} \cdot x^{q - M} ,
\label{eq8c}
\end{equation}

\noindent where $D$ specifies the number of expansion terms, ${\Psi \left( {M,l,q} \right)}$ is a coefficient given by

\begin{equation}
\Psi \left( {M,l,q} \right) = \frac{{{{\left( { - 1} \right)}^q}\sqrt \pi   \cdot \Gamma \left( {2M} \right) \cdot \Gamma \left( {\frac{1}{2} + l - M} \right) \cdot \mathbb{L}\left( {l,q} \right)}}{{{2^{M - q}} \cdot \Gamma \left( {\frac{1}{2} - M} \right) \cdot \Gamma \left( {\frac{1}{2} + l + M} \right) \cdot l!}} 
\label{eq8d}
\end{equation}

\noindent and $\mathbb{L}\left( {l,q} \right) = \left( {\begin{array}{*{20}{c}}{l - 1}\\{q - 1}\end{array}} \right)\frac{{l!}}{{q!}}$ for $\forall l, q > 0$ is the Lah number~\cite{MoluAccess17} with the conventions $\mathbb{L} (0,0) = 1$, $\mathbb{L} (l,0) = 0$, $\mathbb{L} (l,1) = l!$ for $\forall l > 0$.

Since the order $M-1$ of the modified Bessel function of the second kind in \eqref{eq8b} is a non-negative integer number, the series representation \eqref{eq8c} converges except for $M-1 = 0$~\cite{MoluAccess17}. However, $\mathbb{K}_0 (\beta x)$ can be computed from the series representations of $\mathbb{K}_1 (\beta x)$ and $\mathbb{K}_2 (\beta x)$ based on the recurrence identity: ${\mathbb{K}_0}\left( {\beta x} \right) = {\mathbb{K}_2}\left( {\beta x} \right) - 2{\left( {\beta x} \right)^{ - 1}}{\mathbb{K}_1}\left( {\beta x} \right)$~\cite{LinkWolfram}. 

Finally, based on the derivation in the Appendix~C, the tractable closed-form NMSE approximation can be obtained as follows

    \begin{equation}
    NMSE \approx 
       \begin{dcases}
	  		\frac{1}{{\left( {U{N_T} - 1} \right)!}}\sum\limits_{q = 0}^D {{\mathbb{G}_{U{N_T}}}\left( q \right)\left[ {\frac{1}{{{2^{U{N_T} - 1}}}}{\Gamma _{low}}\left( {q + 2,2U{\gamma ^{ - 1/2}}} \right)    } \right.}      \\
	  		\qquad \qquad \qquad    - \frac{1}{{{2^{U{N_T} + 1}}{U^2}{\gamma ^{ - 1}}}}{\Gamma _{low}}\left( {q + 4,2U{\gamma ^{ - 1/2}}} \right)    \\
			\qquad \qquad \qquad    + \ \frac{1}{{{2^{U{N_T} + 3}}{U^4}{\gamma ^{ - 2}}}}{\Gamma _{low}}\left( {q + 6,2U{\gamma ^{ - 1/2}}} \right)     \\
			\qquad \qquad \qquad    + \frac{{{U^2}{\gamma ^{ - 1}}}}{{{2^{U{N_T} - 3}}}}{\Gamma _{up}}\left( {q,2U{\gamma ^{ - 1/2}}} \right)   \\
			\qquad \qquad \qquad    - \frac{{{U^4}{\gamma ^{ - 2}}}}{{{2^{U{N_T} - 5}}}}{\Gamma _{up}}\left. {\left( {q - 2,2U{\gamma ^{ - 1/2}}} \right)} \right],{\rm{  \qquad  \qquad  \qquad    if   \quad  }}U{N_T} > 1   \\
			\sum\limits_{q = 0}^D {{\mathbb{G}_1}\left( q \right)\left[ {{\Gamma _{low}}\left( {q ,2U{\gamma ^{ - 1/2}}} \right) } \right.}    \\
			\qquad \qquad \qquad    - \frac{1}{{4 \cdot {U^2}{\gamma ^{ - 1}}}}{\Gamma _{low}}\left( {q + 2,2U{\gamma ^{ - 1/2}}} \right)   \\
			\qquad \qquad \qquad    + \frac{1}{{16 \cdot {U^4}{\gamma ^{ - 2}}}}{\Gamma _{low}}\left( {q + 4,2U{\gamma ^{ - 1/2}}} \right)   \\
			\qquad \qquad \qquad    + 4 \cdot {U^2}{\gamma ^{ - 1}}{\Gamma _{up}}\left( {q - 2,2U{\gamma ^{ - 1/2}}} \right)  \\
			\qquad \qquad \qquad    - 16 \cdot {U^4}{\gamma ^{ - 2}}{\Gamma _{up}}\left. {\left( {q - 4,2U{\gamma ^{ - 1/2}}} \right)} \right],{\rm{   \qquad  \qquad   \ \ \   if  \quad   }}U{N_T} = 1   
	   \end{dcases}
       \label{eq8e}
    \end{equation}

\noindent in which $\mathbb{G}_{UN_T} (q)$ is calculated as follows

\begin{equation}
\begin{dcases}
{\mathbb{G}_{U{N_T}}}\left( q \right) = \sum\limits_{l = q}^D {\Psi \left( {U{N_T} - 1,l,q} \right)} ,  &  {\rm{  if  \  }}U{N_T} > 1 \\
{\mathbb{G}_1}\left( q \right) = \sum\limits_{l = q}^D {\left( {\Psi \left( {2,l,q} \right) - 2 \cdot \Psi \left( {1,l,q} \right)} \right)}  ,  &  {\rm{  if  \  }}U{N_T} = 1
\end{dcases}
\label{eq8f}
\end{equation}

\section{Asymptotic Analysis}
\label{SecAsymp}

We derive here the asymptotic (in SNRs, BOFs $U$ and the number of antennas $N_T$ parameters) behaviors of the NMSE approximations to gain insights.

\noindent \emph{- \underline{At the intended position:}} 

Considering \eqref{eq6} at high SNR, we have $U{\gamma ^{ - {1 \mathord{\left/ {\vphantom {1 2}} \right. \kern-\nulldelimiterspace} 2}}} \approx 0$, and therefore ${\Gamma _{low}}\left( {s,U{\gamma ^{ - 1/2}}} \right) \approx 0$ and ${\Gamma _{up}}\left( {s,U{\gamma ^{ - 1/2}}} \right) \approx \Gamma (s)$. In \eqref{eq6} the value of the fourth term is much bigger than that of the fifth term. The NMSE at high SNR, $NMSE_{high}$, can be approximated by

\begin{equation}
NMSE_{high} \approx \frac{{{U^2}{\gamma ^{ - 1}}\Gamma \left( {U{N_T} - 2} \right)}}{{\left( {U{N_T} - 1} \right)!}} ,
\label{eq9}
\end{equation}

In the case $a$ is a positive integer, $\Gamma(a) = (a-1)!$. Considering $U N_T > 2$, \eqref{eq9} is further simplified as follows

\begin{equation}
NMS{E_{high}} \approx \frac{{{\gamma ^{ - 1}}}}{{\left( {{N_T} - 1/U} \right)\left( {{N_T} - 2/U} \right)}}  \mathop  \to \limits^{U \to  + \infty } \frac{{{\gamma ^{ - 1}}}}{{{{\left( {{N_T}} \right)}^2}}} .
\label{eq10}
\end{equation}

It reveals that when the BOF is sufficiently high, the NMSE can only be reduced by increasing the number of antennas, as shown later in the simulations. 

Considering \eqref{eq6} at low SNR, the proposed NMSE approximation $NMSE_{low}$ is

\begin{equation}
NMS{E_{low}} \approx \frac{{{\Gamma _{low}}\left( {U{N_T},U{\gamma ^{ - 1/2}}} \right)}}{{\left( {U{N_T} - 1} \right)!}} 
\label{eq11}
\end{equation}

\noindent because in \eqref{eq6} the second and third terms are smaller than the first term, whereas the fourth and fifth terms are approximately equal to 0. Using the series expansion of the lower incomplete Gamma function ${\Gamma _{low}}\left( {a,t} \right) = {t^a}\Gamma \left( a \right){{\rm{e}}^{ - t}}\sum\nolimits_{k = 0}^\infty  {{{{t^k}} \mathord{\left/ {\vphantom {{{t^k}} {\Gamma \left( {a + k + 1} \right)}}} \right. \kern-\nulldelimiterspace} {\Gamma \left( {a + k + 1} \right)}}} $~[20,~eqs.~(8.2.6) and~(8.7.1)] and after some manipulations, we can obtain the following approximation % \cite{LinkGamma}

\begin{equation}
NMS{E_{low}} \approx 1 - {{\rm{e}}^{ - U{\gamma ^{ - 1/2}}}}\sum\limits_{k = 1}^{U{N_T}} {\frac{{{{\left( {U{\gamma ^{ - 1/2}}} \right)}^{U{N_T} - k}}}}{{\left( {U{N_T} - k} \right)!}}} .
\label{eq12}
\end{equation}

It can be seen that the same conclusion as for $NMSE_{high}$ can be drawn from \eqref{eq12}. If the BOF $U$ is sufficiently high, the $NMSE_{low}$ can only be reduced by increasing the number of antennas $N_T$. Because for a fixed value of $N_T$, the term ${{\rm{e}}^{ - U{\gamma ^{ - 1/2}}}}$ reduces faster than the summation term of \eqref{eq12} when increasing $U$, resulting in an un-changed $NMSE_{low}$ value if $U$ is sufficiently high. In the case $U$ is set to a certain value, $NMSE_{low}$ value can still be reduced further when increasing $N_T$ as more terms are added in the summation.

\noindent \emph{- \underline{At the unintended position:}} 

At high SNR, similar to the analysis at the intended position, only the terms with $\Gamma_{up} (\cdot)$ have a significant contribution to the NMSE. Considering $U N_T > 1$ for the sake of simplicity, \eqref{eq8e} can be approximated to% (19). %\eqref{eq8e} can be approximated to \eqref{eq13}. % as follows

    \begin{equation}
    NMS{E_{high}} \approx \frac{{{\gamma ^{ - 1}}\sqrt \pi  {U^2} \cdot \Gamma \left( {2\left( {U{N_T} - 1} \right)} \right)}}{{{2^{2U{N_T} - 4}} \cdot \Gamma \left( {U{N_T}} \right)}}\sum\limits_{q = 0}^D {\sum\limits_{l = q}^D {\frac{{{{\left( { - 2} \right)}^q}\Gamma \left( {\frac{3}{2} - U{N_T} + l} \right) \mathbb{L} \left( {l,q} \right){\Gamma _{up}}\left( {q,2U{\gamma ^{ - 1/2}}} \right)}}{{\Gamma \left( {\frac{3}{2} - U{N_T}} \right)\Gamma \left( {U{N_T} - \frac{1}{2} + q} \right)q!}}} } 
       \label{eq13}
    \end{equation}

Applying the Legendre duplication formula to $\Gamma \left( {2\left( {U{N_T} - 1} \right)} \right)$ and by virtue of $\Gamma \left( {U{N_T}} \right) = \left( {U{N_T} - 1} \right)\Gamma \left( {U{N_T} - 1} \right)$, (19) is rewritten as %in (20). 

    \begin{equation}
    NMS{E_{high}} \approx 2{\gamma ^{ - 1}}\frac{{{U^2}}}{{\left( {U{N_T} - 1} \right)}}\sum\limits_{q = 1}^D {\sum\limits_{l = q}^D {\frac{{{{\left( { - 2} \right)}^q} \mathbb{L} \left( {l,q} \right)\Gamma \left( q \right)\Gamma \left( {U{N_T} - \frac{1}{2}} \right)\Gamma \left( {\frac{3}{2} - U{N_T} + l} \right)}}{{q!\Gamma \left( {\frac{3}{2} - U{N_T}} \right)\Gamma \left( {U{N_T} - \frac{1}{2} + q} \right)}}} }
       \label{eq14}
    \end{equation}

It can be observed that both numerator and denomerator of the fraction term inside the summation of \eqref{eq14} are propotional to $U^2 N_T^2$, therefore this fraction asymptotically converges to a constant when increasing $UN_T$. Considering $U$ and $N_T$ as the variables, $NMSE_{high}$ is propotional to the following expression

\begin{equation}
NMS{E_{high}} \propto \frac{{{U^2}}}{{U{N_T} - 1}}
\label{eq15}
\end{equation}

It can be concluded from \eqref{eq15} that increasing the BOF $U$ causes the increase of NMSE (or equivalently decreases the focusing gain). The NMSE can be reduced by using more transmit antennas. 

The same analysis and conclusion can be drawn for low SNRs, we skip it as it does not bring any additional insight.

\section{Simulation Results}
\label{sec4}

We consider a 256-subcarrier MISO OFDM system, i.e., $Q=256$. A multi-path channel of type Extended Pedestrian A (EPA)~\cite{Chan3GPP} is used in the simulations. Its power delay profile (PDP) is given in Table~\ref{tab1}. The overall channel power is normalized to unity for each channel realization. We assume that the channel is perfectly known at the transmitter. The impact of imperfect channel estimation on the TR-based system performance has been studied in \cite{AlizadehIET12} (and references therein), but is beyond the scope of this paper. Note that, we set the number of the expansion term of modified Bessel function $D=10$ in the following computations. The numerical results at the intended and unintended positions are presented by the solid-lines and dashed-lines, respectively. The analytical results with respect to either intended or unintended positions are plotted with marker symbols.

% TABLE I: Extended Pedestrian A (EPA) PDP
\begin{table}[!htb]
\caption{PDP of the Extended Pedestrian A (EPA) channel.}
\label{tab1}
\centering
% Some packages, such as MDW tools, offer better commands for making tables
% than the plain LaTeX2e tabular which is used here.
\begin{tabular}{|c|c|}
\hline
Excess tap delay (ns) & Relative power (dB) \\
\hline
%\hline
0                                 & 0.0                            \\
\hline
30                               & -1.0                           \\
\hline
70                               & -2.0                           \\
\hline
90                               & -3.0                           \\
\hline
110                             & -8.0                           \\
\hline
190                             & -17.2                         \\
\hline
410                             & -20.8                         \\
\hline
\end{tabular}
\end{table}

\begin{figure*}[!htb]
\centering
\includegraphics[width=1.0\textwidth]{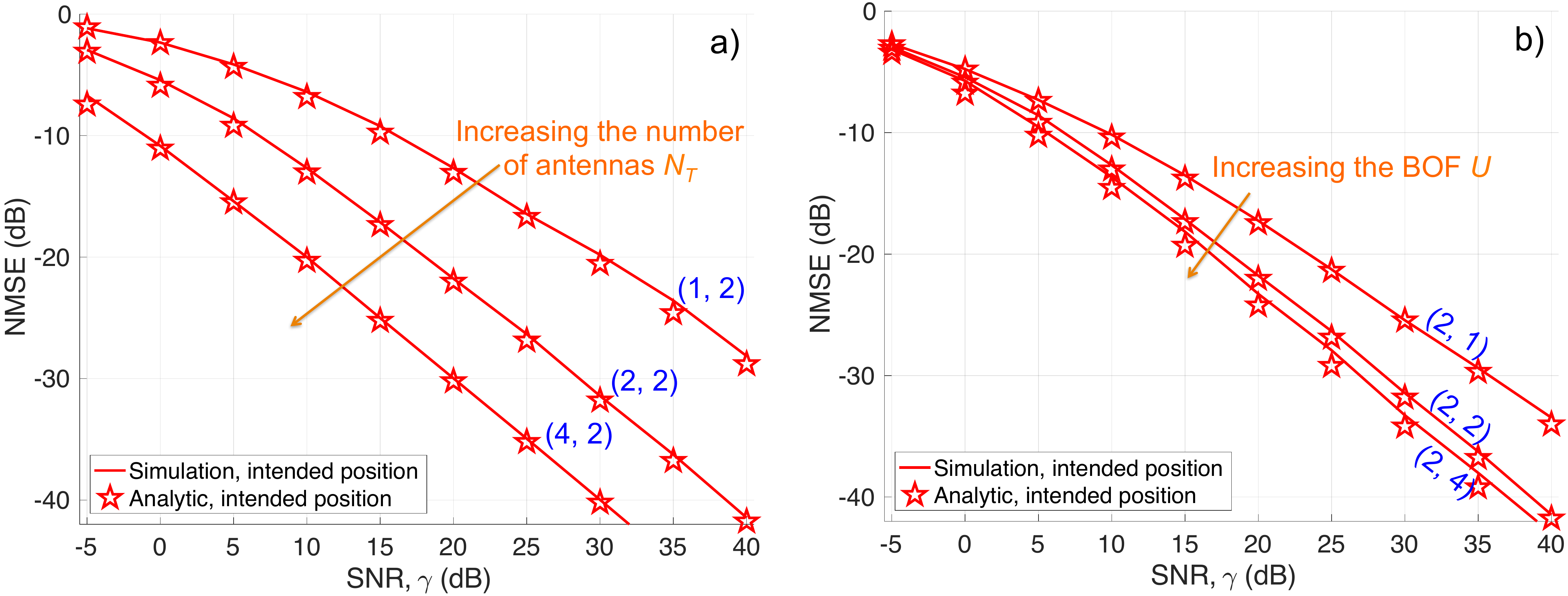}
\caption{NMSE versus SNR at the intended position. a) Different number of antennas when $U = 2$, b) Different BOFs when $N_T = 2$. The couple $(N_T, U)$ is noticed on each curve.}
\label{fig2}
\end{figure*}

\begin{figure*}[!htb]
\centering
\includegraphics[width=1.0\textwidth]{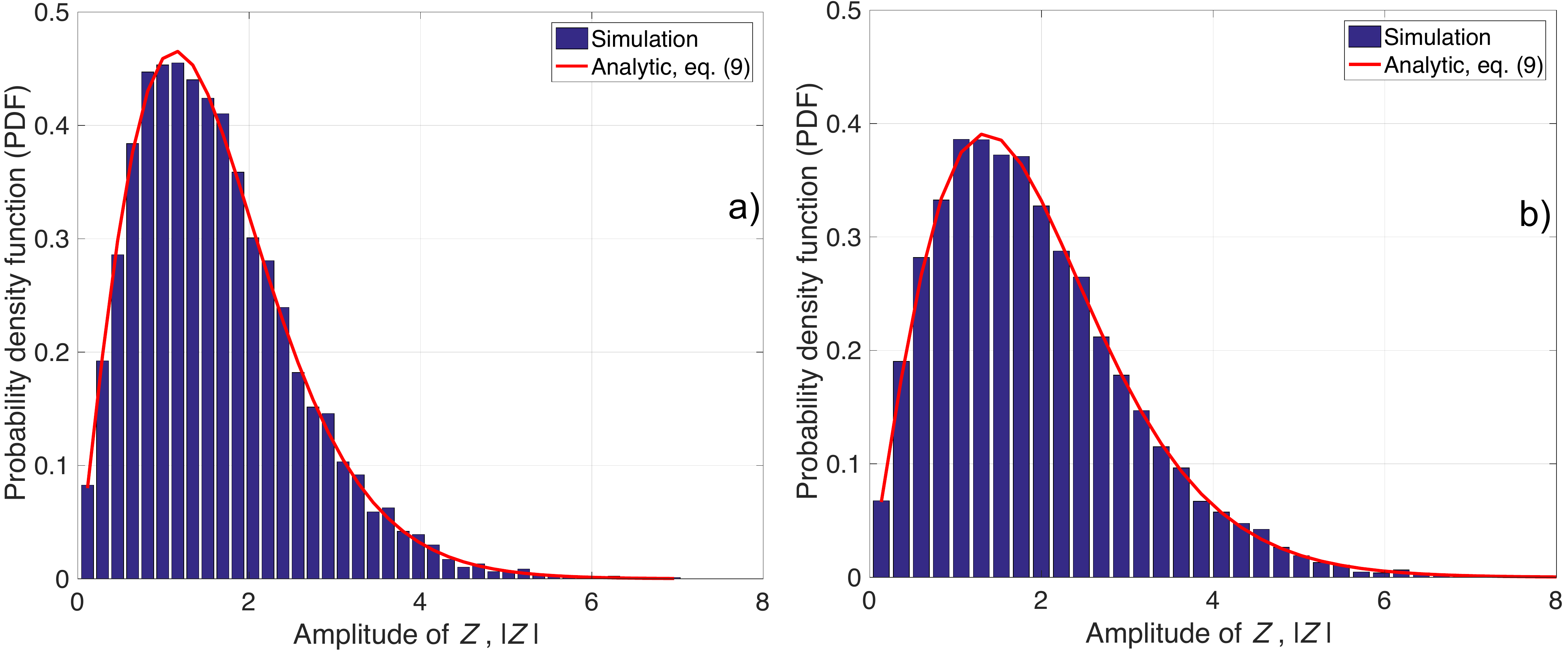} %0.72
\caption{Histogram of simulated RV $\left| Z \right|$ (constructed by the sum of five independent products ($M=5$) of 2 complex zero-mean RVs) and the associated PDF envelopes. a) $\sigma_{Y_1} = 0.6$ and $\sigma_{Y_2} = 1.4$; b) $\sigma_{Y_1} = \sigma_{Y_2} = 1$.}
\label{fig3}
\end{figure*}

In the first step, we evaluate the NMSE as a function of the SNR at the intended position in Fig.~\ref{fig2}. When $U$ is equal to 2, Fig.~2(a) shows NMSE evolution for different number of antennas. When $N_T$ is set to 2, Fig.~2(b) presents NMSE evolution for different BOFs. 
As expected, the analytical closed-form NMSEs match the ones obtained by simulations, confirming the correctness of our derivation in \eqref{eq6}. The simulation results also confirm our previous observation that when increasing the BOF, the NMSE (and hence the spatio-temporal focusing) converges asymptotically to an NMSE lower bound. It is observed that increasing the number of antennas continuously provides the focusing gain. For instance, in order to obtain a NMSE = -20~dB, by changing the number of antennas from one to two, we achieve a SNR gain of about 12 dB, while by using a BOF of two instead of one, we can only obtain a SNR gain of about 6 dB.

\begin{figure*}[!htb]
\centering
\includegraphics[width=1.0\textwidth]{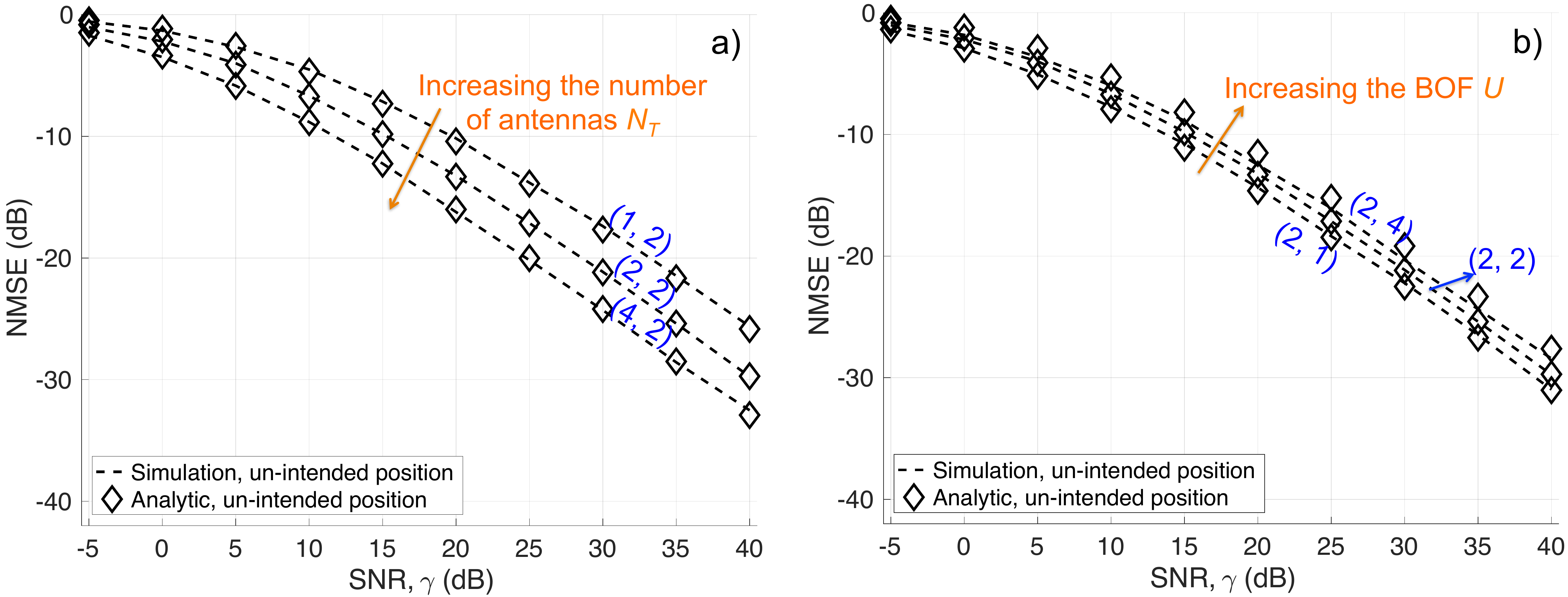} % 0.4
\caption{NMSE versus SNR at the unintended position. a) Different number of antennas when $BOF = 2$, b) Different BOFs when $N_T = 2$. The couple $(N_T, U)$ is noticed on each curve.}
\label{fig4}
\end{figure*}

Before investigating the MSE derivation at the unintended position, we verify the correctness of the derived PDF formula \eqref{eq8pdf}. 
Fig.~\ref{fig3} presents the examples of the histograms of simulated RV $\left| Z \right|$ and its PDF envelope when $M=5$. Particularly, Fig.~\ref{fig3}(a) shows the histogram of $\left| Z \right|$ and the associated PDF envelope using \eqref{eq8pdf} when $\sigma_{Y_1} = 0.6$ and $\sigma_{Y_2} = 1.4$, whereas Fig.~\ref{fig3}(b) illustrates the histogram of $\left| Z \right|$ and the associated PDF envelope applying \eqref{eq8pdf} when the variances are set to unities. It can be seen that our derived PDF fits the simulation histogram, which, along with the detailed derivation in Appendix~B, highlights our contribution.

We now evaluate the NMSE as a function of the SNR at the unintended position in Fig.~\ref{fig4} with different number of antennas when $U$ is equal to 2 (Fig.~4(a)) and with different BOFs when $N_T$ is set to 2 (Fig.~4(b)). 

Similar to the results at the intended position, the numerical NMSEs follow the predicted analytical ones in \eqref{eq8e}. The simulation results also confirm the asymptotic analyses that the NMSE can only improve by increasing the number of antennas, while, in contrast to the case at the intended position, increasing the BOF gives a poorer MSE. For instance, at NMSE = -20~dB, by changing the number of antennas from one to two, we achieve about 4 dB SNR gain (much less than that at the intended position), while by using a BOF of two instead of one, the SNR loss is about 0.6 dB.

\begin{figure}%[!t]
\centering
\includegraphics[width=0.52\textwidth]{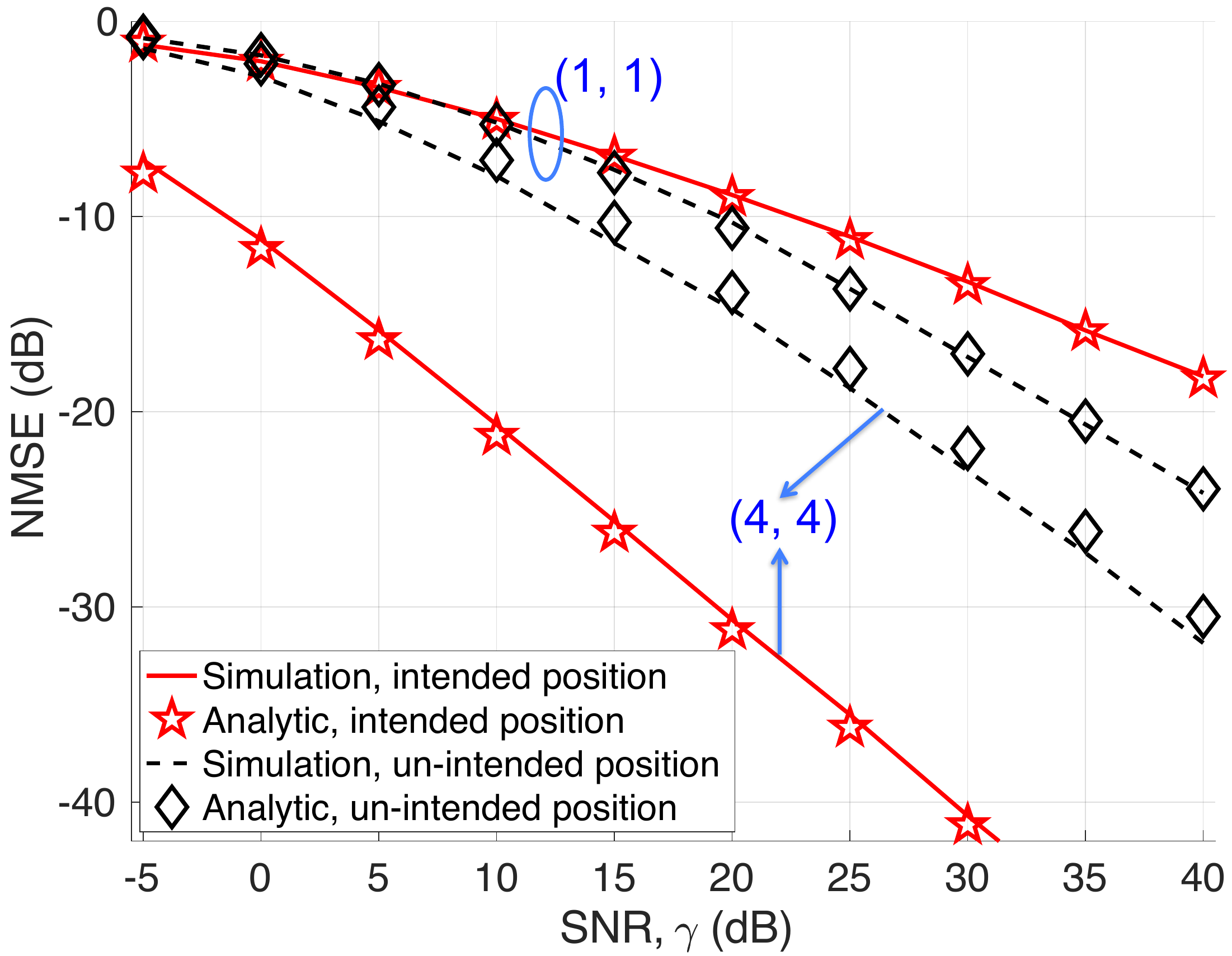}
\caption{NMSE versus SNR at both intended and unintended positions when increasing the number of antennas and BOFs. The couple $(N_T, U)$ is noticed on each curve.}
\label{fig5}
\end{figure}

The MSEs at the intended and unintended positions are compared in Fig.~\ref{fig5}. It can be observed that when changing from the no-focusing-gain case ($U = N_T = 1$) to the case where $U = N_T = 4$ and considering the NMSE = -20~dB, the SNR gain at the intended position is about 22~dB bigger than that at the unintended position, confirming again the focusing effect provided by the TR precoding. It is also reminded that the SNR gain at the unintended position is provided by only the increase of the number of antennas. Surprisingly, we observe from the no-focusing-gain case that the NMSE value at the unintended position is smaller than that at the intended position for the moderate and high SNRs (Fig.~\ref{fig5}). Rigorously, we can explain this fact based on the equations \eqref{eq6} and \eqref{eq8e} by considering their monotonicities in the interesting SNR range. Intuitively, it can be explained because the precoded channels at the unintended position are complex, while the precoded channels at the intended position are real, resulting in a less deep fading probability at the subcarrier-level of the equivalent channel magnitude.

\begin{figure}%[!t]
\centering
\includegraphics[width=0.52\textwidth]{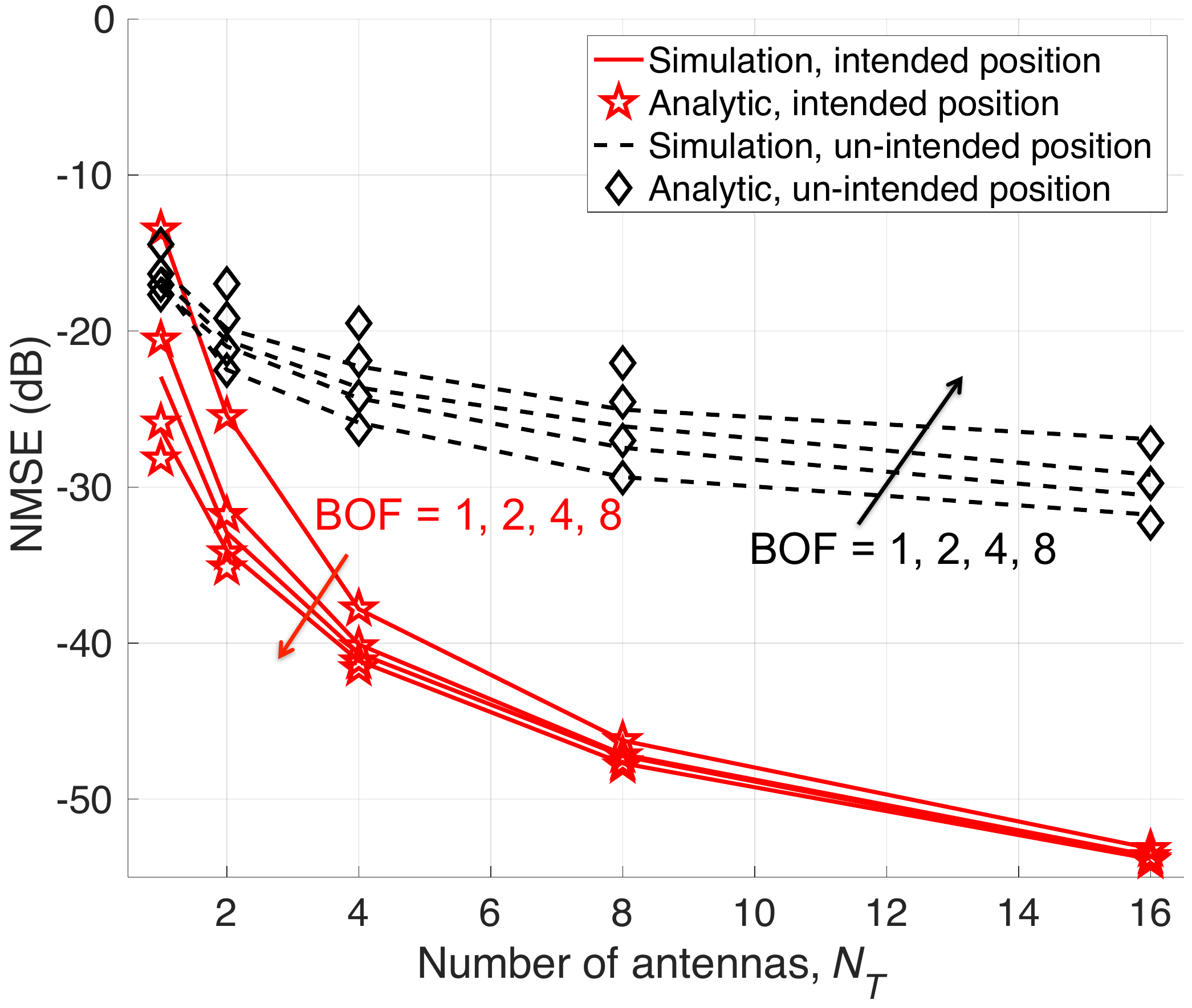}
\caption{NMSE of received symbols when using different number of antennas and BOFs. SNR = 30 dB.}
\label{fig6}
\end{figure}

Finally, we set the SNR to 30~dB. We investigate the NMSE values when varying the number of antennas and BOFs. The results are presented in Fig.~\ref{fig6}. At the intended position, as suggested in the asymptotic analysis in Section~\ref{sec3}, the NMSE does not improve further when the BOF is sufficiently high. In this case, we can only improve the NMSE by increasing the number of antennas. At the unintended position, although the NMSE can be reduced slowly by using more transmit antennas, the NMSE worsens when BOF changes from 1 to 8. The numerical results confirm again the observation made in the analyses.  It should be reminded that when the BOF value is high, the assumption of the statistical independence among RVs $H_n$ does not hold so that there are some mismatches between the analytical and numerical results.

The focusing gain is defined by the SNR gain necessary to keep the MSE at a fixed value. At the intended position, the RV $K_n$ is built constructively thanks to the TR precoding of the signal with the corresponding CFR. This leads to the reduction of the NMSE and hence improves the focusing gain, especially when increasing BOFs (associated with the frequency diversity gain). On the contrary, at the unintended position, the RV $K_n$ is constructed destructively from $(H_n^k)^*$ and $\overline{H} _n^k$ associated with the CFR component at the unintended position, because of their spatial independence. This causes the increase in the NMSE in \eqref{eq5} when increasing the BOFs. In the case we increase the number of antennas, the NMSEs in both intended and unintended positions are reduced thanks to the spatial diversity gain. However, the SNR gain at the unintended position is smaller than that at the intended position, since the spatial diversity gain at the unintended position can partially compensates for the destructive construction of the RV $K_n$ (originating from the BOF).

\section{Conclusion}

We have presented a proper way to perform FD/TR precoding in MISO OFDM communication systems for BOFs different than one. We have also verified the focusing gain (provided by the FD/TR precoding), which is defined as the SNR gain required to maintain a fixed value of MSE. By evaluating the NMSE, we have shown that, at the intended position, increasing either the BOF or the number of antennas improves the focusing gain. Conversely, at the unintended position, the useful received power is lower. The closed-form NMSE approximations at the intended and unintended positions have been derived and validated through simulations.

\section*{Appendix A}
\label{AppenA}
% Derivation of the NMSE approximation @ the intended position

At the intended position, based on the PDF of $K_n$, \eqref{eq5a} can be rewritten as

\begin{align}
NMSE & = \int\limits_0^\infty  {\frac{{{\gamma ^{ - 1}}}}{{{{{z^2}} \mathord{\left/
 {\vphantom {{{z^2}} {{U^2}}}} \right.
 \kern-\nulldelimiterspace} {{U^2}}} + {\gamma ^{ - 1}}}}\frac{{{z^{M - 1}}}}{{\left( {M - 1} \right)!}}{{\rm{e}}^{ - z}}dz}      \nonumber \\
           & = \frac{1}{{\left( {M - 1} \right)!}}\left( {\int\limits_0^{U{\gamma ^{ - 1/2}}} {\frac{{{\gamma ^{ - 1}}}}{{{{{z^2}} \mathord{\left/
 {\vphantom {{{z^2}} {{U^2}}}} \right. \kern-\nulldelimiterspace} {{U^2}}} + {\gamma ^{ - 1}}}}{z^{M - 1}}{{\rm{e}}^{ - z}}dz}  } \right.     {\rm{   }} + \left. {\int\limits_{U{\gamma ^{ - 1/2}}}^\infty  {\frac{{{\gamma ^{ - 1}}}}{{{{{z^2}} \mathord{\left/
 {\vphantom {{{z^2}} {{U^2}}}} \right. \kern-\nulldelimiterspace} {{U^2}}} + {\gamma ^{ - 1}}}}{z^{M - 1}}{{\rm{e}}^{ - z}}dz} } \right) .
\label{eqB1}
\end{align}

Splitting the NMSE formula into two parts ${T_1} = \int_0^{U{\gamma ^{ - 1/2}}} {\frac{{{\gamma ^{ - 1}}}}{{{{{z^2}} \mathord{\left/ {\vphantom {{{z^2}} {{U^2}}}} \right. \kern-\nulldelimiterspace} {{U^2}}} + {\gamma ^{ - 1}}}}{z^{M - 1}}{{\rm{e}}^{ - z}}dz} $ and ${T_2} = \int_{U{\gamma ^{ - 1/2}}}^\infty  {\frac{{{\gamma ^{ - 1}}}}{{{{{z^2}} \mathord{\left/ {\vphantom {{{z^2}} {{U^2}}}} \right. \kern-\nulldelimiterspace} {{U^2}}} + {\gamma ^{ - 1}}}}{z^{M - 1}}{{\rm{e}}^{ - z}}dz} $ ensures that the integrals can be simplified in the two ranges of interest. More particularly, $T_1$ can be rewritten as

\begin{align}
{T_1} & = \int\limits_0^{U{\gamma ^{ - 1/2}}} {{{\left( {1 + \frac{{{z^2}}}{{{U^2}{\gamma ^{ - 1}}}}} \right)}^{ - 1}}{z^{M - 1}}{{\rm{e}}^{ - z}}dz}     \nonumber \\
          &  \approx \int\limits_0^{U{\gamma ^{ - 1/2}}} {\left( {1 - \frac{{{z^2}}}{{{U^2}{\gamma ^{ - 1}}}} + \frac{{{z^4}}}{{{U^4}{\gamma ^{ - 2}}}}} \right){z^{M - 1}}{{\rm{e}}^{ - z}}dz} ,
\label{eqB2}
\end{align}

\noindent where the approximation is achieved by using the Taylor expansion ${\left( {1 + x} \right)^{ - 1}} = \sum\nolimits_{n = 0}^\infty  {{{\left( { - x} \right)}^n}} $ that converges on the range of the integral $\left( {0,U{\gamma ^{ - 1/2}}} \right)$, as ${{{z^2}} \mathord{\left/ {\vphantom {{{z^2}} {\left( {{U^2}{\gamma ^{ - 1}}} \right)}}} \right. \kern-\nulldelimiterspace} {\left( {{U^2}{\gamma ^{ - 1}}} \right)}} < 1$. Substituting the lower incomplete Gamma function defined in \eqref{eq7} into \eqref{eqB2}, we obtain the closed-form expression of $T_1$

\begin{align}
{T_1} \approx & \ {\Gamma _{low}}\left( {M,U{\gamma ^{ - 1/2}}} \right) - \frac{1}{{{U^2}{\gamma ^{ - 1}}}}{\Gamma _{low}}\left( {M + 2,U{\gamma ^{ - 1/2}}} \right)   \nonumber \\
                       & + \frac{1}{{{U^4}{\gamma ^{ - 2}}}}{\Gamma _{low}}\left( {M + 4,U{\gamma ^{ - 1/2}}} \right) .
\label{eqB3}
\end{align}

Due to the fact that ${{{U^2}{\gamma ^{ - 1}}} \mathord{\left/ {\vphantom {{{U^2}{\gamma ^{ - 1}}} {{z^2}}}} \right. \kern-\nulldelimiterspace} {{z^2}}} < 1$, we can apply again the Taylor expansion to $T_2$ that ensures the convergence on the range of the integral $\left( {U{\gamma ^{ - 1/2}},\infty } \right)$, $T_2$ can be rewritten as follows

\begin{align}
{T_2} & = \int\limits_{U{\gamma ^{ - 1/2}}}^\infty  {{U^2}{\gamma ^{ - 1}}{z^{ - 2}}{{\left( {1 + \frac{{{U^2}{\gamma ^{ - 1}}}}{{{z^2}}}} \right)}^{ - 1}}{z^{M - 1}}{{\rm{e}}^{ - z}}dz}     \nonumber \\
          &  \approx \int\limits_{U{\gamma ^{ - 1/2}}}^\infty  {{U^2}{\gamma ^{ - 1}}{z^{ - 2}}\left( {1 - \frac{{{U^2}{\gamma ^{ - 1}}}}{{{z^2}}}} \right){z^{M - 1}}{{\rm{e}}^{ - z}}dz} .
\label{eqB4}
\end{align}

Using the definition of the upper incomplete Gamma function \eqref{eq8}, the closed-form expression of $T_2$ can be derived

\begin{align}
{T_2} = & \ {U^2}{\gamma ^{ - 1}}{\Gamma _{up}}\left( {M - 2,U{\gamma ^{ - 1/2}}} \right)     - {U^4}{\gamma ^{ - 2}}{\Gamma _{up}}\left( {M - 4,U{\gamma ^{ - 1/2}}} \right) .
\label{eqB5}
\end{align}

Finally, substituting \eqref{eqB3} and \eqref{eqB5} into \eqref{eqB1} and using the fact that $M=U N_T$, we achieve the closed-form NMSE expression at the intended position as in \eqref{eq6}.% $\square$.

\section*{Appendix B}
\label{AppenB}
% Derivation of the PDF of modulus of sum of products of two complex Gaussian RVs

Considering a RV $Z_m = Y_{1,m} \cdot Y_{2,m}$, where $Y_{1,m} \sim \mathcal{CN}(0, \,\sigma_{Y_1}^{2})\,$, $Y_{2,m} \sim \mathcal{CN}(0, \,\sigma_{Y_2}^{2})\,$, and defining $R_m = \left| Z_m \right|$, $\Theta_m = \angle Z_m$, where $\angle$ is the angle operator, the marginal joint PDF of $R_m$ and $\Theta_m$ has been derived in~[23,~eq.~(17)] as follows %~\cite{ODonoughueTSP12}

\begin{equation}
{f_{{R_m},{\Theta _m}}}\left( {{r_m},{\theta _m}} \right) = \frac{{2{r_m}}}{{\pi \sigma _{Y_1}^2\sigma _{Y_2}^2}}{\mathbb{K}_0}\left( {\frac{{2r_m}}{{{\sigma _{Y_1}}{\sigma _{Y_2}}}}} \right) .
\label{eqC1}
\end{equation}

\noindent where $\mathbb{K}_M(\cdot)$ is the $M$-th order modified Bessel function of the second kind. The characteristic function (CF) of ${f_{{R_m},{\Theta _m}}}\left( {{r_m},{\theta _m}} \right)$ can be written as
% after the Catersian-polar transformation 

\begin{align}
{\psi _{{R_m},{\Theta _m}}}\left( {j{\omega _1},j{\omega _2}} \right) & = \int\limits_{ - \infty }^\infty  {\int\limits_{ - \infty }^\infty  {{{\rm{e}}^{\left( {j{\omega _1}{r_m}\cos {\theta _m} + j{\omega _2}{r_m}\sin {\theta _m}} \right)}} } }   \cdot {f_{{R_m},{\Theta _m}}}\left( {{r_m},{\theta _m}} \right)d{r_m}d{\theta _m}     \nonumber \\
&   = \int\limits_{ 0 }^\infty  {\int\limits_0^{2\pi } {{{\rm{e}}^{\left( {j{\omega _1}{r_m}\cos {\theta _m} + j{\omega _2}{r_m}\sin {\theta _m}} \right)}}d{\theta _m}  } }    \cdot \frac{2}{{\pi \sigma _{Y_1}^2\sigma _{Y_2}^2}}{r_m}{\mathbb{K}_0}\left( {\frac{{2{r_m}}}{{{\sigma _{Y_1}}{\sigma _{Y_2}}}}} \right)d{r_m} 
\label{eqC2}
\end{align}

Applying the equation [24,~eq.~(3.937-2)] to the inner integral of \eqref{eqC2}, we can obtain the following result % in~\cite{GradshteynBook07}

\begin{equation}
\int\limits_0^{2\pi } {{{\rm{e}}^{\left( {j{\omega _1}{r_m}\cos {\theta _m} + j{\omega _2}{r_m}\sin {\theta _m}} \right)}}d{\theta _m}}  = 2\pi {\mathbb{I}_0}\left( {j{r_m}\sqrt {\omega _1^2 + \omega _2^2} } \right)
\label{eqC3}
\end{equation}

\noindent where $\mathbb{I}_M (\cdot)$ is the modified Bessel function of the first kind with order $M$. The CF of of ${f_{{R_m},{\Theta _m}}}\left( {{r_m},{\theta _m}} \right)$ can be rewritten as

\begin{align}
{\psi _{{R_m},{\Theta _m}}}\left( {j{\omega _1},j{\omega _2}} \right) = \frac{4}{{\sigma _{Y_1}^2\sigma _{Y_2}^2}} \cdot        {\rm{   }}\int\limits_0^\infty  {{r_m} \cdot {\mathbb{I}_0}\left( {j{r_m}\sqrt {\omega _1^2 + \omega _2^2} } \right) \cdot {\mathbb{K}_0}\left( {\frac{{2{r_m}}}{{{\sigma _{Y_1}}{\sigma _{Y_2}}}}} \right)d{r_m}}
\label{eqC4}
\end{align}

Using the integration formula [24,~eq.~(6.576-3)], we achieve the CF as follows %  in~\cite{GradshteynBook07}

\begin{equation}
{\psi _{{R_m},{\Theta _m}}}\left( {j{\omega _1},j{\omega _2}} \right) = {}_2{F_1}\left( {1,1;1; - \frac{{\sigma _{Y_1}^2\sigma _{Y_2}^2\left( {\omega _1^2 + \omega _2^2} \right)}}{4}} \right) ,
\label{eqC5}
\end{equation}

\noindent where ${}_2{F_1}\left( {\alpha ,\beta;\lambda;z} \right)$ is the Gaussian hypergeometric function. Due to the fact that ${}_2{F_1}\left( {\alpha ,1;1;z} \right) = {\left( {1 - z} \right)^{ - \alpha }}$, \eqref{eqC5} becomes

\begin{equation}
{\psi _{{R_m},{\Theta _m}}}\left( {j{\omega _1},j{\omega _2}} \right) = {\left( {1 + \frac{{\sigma _{Y_1}^2\sigma _{Y_2}^2\left( {\omega _1^2 + \omega _2^2} \right)}}{4}} \right)^{ - 1}}.
\label{eqC6}
\end{equation}

Considering the RV $Z = \sum\nolimits_{m = 0}^{M - 1} {{Z_m}} $ with its modulus $R = \left| Z \right|$ and its argument $\Theta = \angle Z$ and assuming that $Z_m$ and $Z_n$ are statistically independent for $\forall m \ne n$, the CF of ${f_{R,\Theta }}\left( {r,\theta } \right)$ corresponding to the RV $Z$ can be derived based on the properties of CF as follows

\begin{equation}
{\psi _{R,\Theta }}\left( {j{\omega _1},j{\omega _2}} \right) = {\left( {1 + \frac{{\sigma _{Y_1}^2\sigma _{Y_2}^2\left( {\omega _1^2 + \omega _2^2} \right)}}{4}} \right)^{ - M}} .
\label{eqC7}
\end{equation}

Inverting the CF ${\psi _{R,\Theta }}\left( {j{\omega _1},j{\omega _2}} \right)$, we can obtain the PDF of $Z$. Instead of directly inverting this CF, which is a non-trivial problem, we consider the following joint PDF

\begin{equation}
{f_{R,\Theta }}\left( {r,\theta } \right) = \frac{{2{r^M}}}{{\pi  \cdot \Gamma \left( M \right) \cdot {{\left( {{\sigma _{Y_1}}{\sigma _{Y_2}}} \right)}^{M + 1}}}}{\mathbb{K}_{M - 1}}\left( {\frac{{2r}}{{{\sigma _{Y_1}}{\sigma _{Y_2}}}}} \right) .
\label{eqC8}
\end{equation}

After carrying out the same manipulations as for ${f_{{R_m},{\Theta _m}}}\left( {{r_m},{\theta _m}} \right)$ (from \eqref{eqC2} to \eqref{eqC6}), we get the same CF as in \eqref{eqC7}. Therefore, we can conclude that ${f_{R,\Theta }}\left( {r,\theta } \right)$ function in \eqref{eqC8} is the marginal joint PDF of the RV $Z$. Integrating the joint PDF over the RV $\theta$, we obtain the PDF of $\left| Z \right|$ as in~\eqref{eq8pdf}.

\section*{Appendix C}
\label{AppenC}
% Derivation of the NMSE approximation @ the unintended position

From \eqref{eq8c} and \eqref{eq8f}, $\mathbb{K}_{M-1} (2z)$ in \eqref{eq8b} can be approximated by

\begin{equation}
{\mathbb{K}_{M - 1}}\left( {2z} \right) \approx 
\begin{cases}
{e^{ - 2z}}\sum\limits_{q = 0}^D {{\mathbb{G}_M}\left( q \right) \cdot {{\left( {2z} \right)}^{q - M + 1}}},  & {\rm{ \ \  if  \ }}M > 1   \\
{e^{ - 2z}}\sum\limits_{q = 0}^D {{\mathbb{G}_1}\left( q \right) \cdot {{\left( {2z} \right)}^{q - 2}}},          & {\rm{ \ \  if  \ }}M = 1
\end{cases}
\label{eqD1}
\end{equation}

Similar to the derivation in the Appendix~A, we split the NMSE at the unintended position \eqref{eq8b} into two integrals to ensure the convergence when applying the Taylor expansion. More particularly, \eqref{eq8b} can be rewritten as

\begin{align}
NMSE = & \ \frac{4}{{\Gamma \left( M \right)}}\int\limits_0^{U{\gamma ^{ - 1/2}}} {{{\left( {1 + \frac{{{z^2}}}{{{U^2}{\gamma ^{ - 1}}}}} \right)}^{ - 1}}{z^M}{\mathbb{K}_{M - 1}}\left( {2z} \right)dz}   \nonumber  \\
              & {\rm{  \ }} + \frac{4}{{\Gamma \left( M \right)}}\int\limits_{U{\gamma ^{ - 1/2}}}^\infty  {{U^2}{\gamma ^{ - 1}}{{\left( {1 + \frac{{{U^2}{\gamma ^{ - 1}}}}{{{z^2}}}} \right)}^{ - 1}}{z^{M - 2}}{\mathbb{K}_{M - 1}}\left( {2z} \right)dz}  .
\label{eqD2}
\end{align}

Applying the Taylor series expansion to \eqref{eqD2}, we obtain the following NMSE approximation

\begin{align}
NMSE \approx \ & \frac{4}{{\Gamma \left( M \right)}}\int\limits_0^{U{\gamma ^{ - 1/2}}} {{z^M}{\mathbb{K}_{M - 1}}\left( {2z} \right)dz}   - \frac{4}{{\Gamma \left( M \right){U^2}{\gamma ^{ - 1}}}}\int\limits_0^{U{\gamma ^{ - 1/2}}} {{z^{M + 2}}{\mathbb{K}_{M - 1}}\left( {2z} \right)dz}    \nonumber \\
   \ \   & + \frac{4}{{\Gamma \left( M \right){U^4}{\gamma ^{ - 2}}}}\int\limits_0^{U{\gamma ^{ - 1/2}}} {{z^{M + 4}}{\mathbb{K}_{M - 1}}\left( {2z} \right)dz}   + \frac{{4{U^2}{\gamma ^{ - 1}}}}{{\Gamma \left( M \right)}}\int\limits_{U{\gamma ^{ - 1/2}}}^\infty  {{z^{M - 2}}{\mathbb{K}_{M - 1}}\left( {2z} \right)dz}   \nonumber \\
    \ \   & - \frac{{4{U^4}{\gamma ^{ - 2}}}}{{\Gamma \left( M \right)}}\int\limits_{U{\gamma ^{ - 1/2}}}^\infty  {{z^{M - 4}}{\mathbb{K}_{M - 1}}\left( {2z} \right)dz} .
\label{eqD3}
\end{align}

We derive the first integral in~\eqref{eqD3} using the approximation in~\eqref{eqD1} as follows

\begin{align}
J  _1 & = \frac{4}{{\Gamma \left( M \right)}}\int\limits_0^{U{\gamma ^{ - 1/2}}} {{z^M}{\mathbb{K}_{M - 1}}\left( {2z} \right)dz}  \nonumber  & \\
         & \approx \frac{4}{{\Gamma \left( M \right){2^M}}} 
\begin{cases}
\int\limits_0^{U{\gamma ^{ - 1/2}}} {\sum\limits_{q = 0}^D {{\mathbb{G}_M}\left( q \right) \cdot {{\left( {2z} \right)}^{q + 1}} \cdot {{\rm{e}}^{ - 2z}}} dz} ,   {\rm{ \qquad  if \ \  }}M > 1   \\
\int\limits_0^{U{\gamma ^{ - 1/2}}} {\sum\limits_{q = 0}^D {{\mathbb{G}_1}\left( q \right) \cdot {{\left( {2z} \right)}^{q-1}} \cdot {{\rm{e}}^{ - 2z}}} dz} ,   {\rm{ \ \qquad  if \ \  }}M = 1   \\
\end{cases}
\label{eqD4}
\end{align}

By changing the variable, i.e., $x = 2z$, after some manipulations, $J_1$ can be derived as follows

\begin{align}
J_1 \approx
\begin{cases}
\frac{1}{{\left( {M - 1} \right)!{2^{M - 1}}}}\sum\limits_{q = 0}^D {{\mathbb{G}_M}\left( q \right) \cdot {\Gamma _{low}}\left( {q + 2,2U{\gamma ^{ - 1/2}}} \right)} ,   {\rm{ \qquad if \ \  }}M > 1   \\
\sum\limits_{q = 0}^D {{\mathbb{G}_1}\left( q \right) \cdot {\Gamma _{low}}\left( {q ,2U{\gamma ^{ - 1/2}}} \right)} ,  {\rm{ \qquad \qquad \qquad \qquad \quad    if \ \  }}M = 1
\end{cases}
\label{eqD5}
\end{align}

Carrying out the same derivation for the four integrals left in~\eqref{eqD3}, we obtain the following results

\begin{align}
{J_2} & = \frac{4}{{\Gamma \left( M \right){U^2}{\gamma ^{ - 1}}}}\int\limits_0^{U{\gamma ^{ - 1/2}}} {{z^{M + 2}}{\mathbb{K}_{M - 1}}\left( {2z} \right)dz}  \nonumber \\
         & \approx 
         \begin{dcases}
         \frac{1}{{{U^2}{\gamma ^{ - 1}}\left( {M - 1} \right)!{2^{M + 1}}}} \cdot   \sum\limits_{q = 0}^D {{\mathbb{G}_M}\left( q \right) \cdot {\Gamma _{low}}\left( {q + 4,2U{\gamma ^{ - 1/2}}} \right)} ,{\rm{ \quad  if \ \ }}M > 1   \\
         \frac{1}{{4{U^2}{\gamma ^{ - 1}}}}\sum\limits_{q = 0}^D {{\mathbb{G}_1}\left( q \right) \cdot {\Gamma _{low}}\left( {q + 2,2U{\gamma ^{ - 1/2}}} \right)} ,   {\rm{ \qquad \qquad \qquad \qquad   if \ \  }}M = 1
         \end{dcases}
\label{eqD6}
\end{align}

\begin{align}
{J_3} & = \frac{4}{{\Gamma \left( M \right){U^4}{\gamma ^{ - 2}}}}\int\limits_0^{U{\gamma ^{ - 1/2}}} {{z^{M + 4}}{\mathbb{K}_{M - 1}}\left( {2z} \right)dz}  \nonumber \\
         & \approx 
         \begin{dcases}
         \frac{1}{{{U^4}{\gamma ^{ - 2}}\left( {M - 1} \right)!{2^{M + 3}}}} \cdot   \sum\limits_{q = 0}^D {{\mathbb{G}_M}\left( q \right) \cdot {\Gamma _{low}}\left( {q + 6,2U{\gamma ^{ - 1/2}}} \right)} ,{\rm{ \ \  if \ \ }}M > 1   \\
         \frac{1}{{16{U^4}{\gamma ^{ - 2}}}}\sum\limits_{q = 0}^D {{\mathbb{G}_1}\left( q \right) \cdot {\Gamma _{low}}\left( {q + 4,2U{\gamma ^{ - 1/2}}} \right)} ,   {\rm{ \qquad \qquad \qquad \quad  if \ \  }}M = 1
         \end{dcases}
\label{eqD7}
\end{align}

\begin{align}
{J_4} & = \frac{{4{U^2}{\gamma ^{ - 1}}}}{{\Gamma \left( M \right)}}\int\limits_{U{\gamma ^{ - 1/2}}}^\infty  {{z^{M - 2}}{\mathbb{K}_{M - 1}}\left( {2z} \right)dz}  \nonumber \\
         & \approx 
         \begin{dcases}
         \frac{{{U^2}{\gamma ^{ - 1}}}}{{\left( {M - 1} \right)!{2^{M - 3}}}}\sum\limits_{q = 0}^D {{\mathbb{G}_M}\left( q \right) \cdot {\Gamma _{up}}\left( {q,2U{\gamma ^{ - 1/2}}} \right)} ,   {\rm{  \qquad \qquad  \quad if \ \  }}M > 1   \\
         4{U^2}{\gamma ^{ - 1}}\sum\limits_{q = 0}^D {{\mathbb{G}_1}\left( q \right) \cdot {\Gamma _{up}}\left( {q - 2,2U{\gamma ^{ - 1/2}}} \right)} ,   {\rm{  \qquad \qquad  \qquad \quad  if \ \  }}M = 1
         \end{dcases}
\label{eqD8}
\end{align}

\noindent and

\begin{align}
{J_5} & = \frac{{4{U^4}{\gamma ^{ - 2}}}}{{\Gamma \left( M \right)}}\int\limits_{U{\gamma ^{ - 1/2}}}^\infty  {{z^{M - 4}}{\mathbb{K}_{M - 1}}\left( {2z} \right)dz}  \nonumber \\
         & \approx 
         \begin{dcases}
         \frac{{{U^4}{\gamma ^{ - 2}}}}{{\left( {M - 1} \right)!{2^{M - 5}}}}\sum\limits_{q = 0}^D {{\mathbb{G}_M}\left( q \right) \cdot {\Gamma _{up}}\left( {q-2,2U{\gamma ^{ - 1/2}}} \right)} ,   {\rm{ \qquad \qquad if \ \  }}M > 1   \\
         16{U^4}{\gamma ^{ - 2}}\sum\limits_{q = 0}^D {{\mathbb{G}_1}\left( q \right) \cdot {\Gamma _{up}}\left( {q - 4,2U{\gamma ^{ - 1/2}}} \right)} ,    {\rm{ \qquad \qquad \qquad \quad  if \ \  }}M = 1
         \end{dcases}
\label{eqD9}
\end{align}

Due to the fact that $M = UN_T$, $\Gamma(M) = (M-1)!$ for a positive integer $M$, substituting~\eqref{eqD5}, \eqref{eqD6}, \eqref{eqD7}, \eqref{eqD8} and \eqref{eqD9} into \eqref{eqD3} yields the closed-form NMSE approximation as in~\eqref{eq8e}.

% use section* for acknowledgment
\section*{Acknowledgment}

The authors would like to thank the financial support of the Copine-IoT Innoviris project, the Icity.Brussels project and the FEDER/EFRO grant.

\end{document}